\documentclass[amsmath,amssymb,superscriptaddress,nobalancelastpage,prb,twocolumn]{revtex4-2} %, linenumbers

\AtBeginDocument{%
\newwrite\bibnotes
\def\bibnotesext{Notes.bib}
\immediate\openout\bibnotes=\jobname\bibnotesext
\immediate\write\bibnotes{@CONTROL{REVTEX41Control}}
\immediate\write\bibnotes{@CONTROL{%
apsrev41Control,author="08",editor="1",pages="1",title="0",year="1"}}
\if@filesw
\immediate\write\@auxout{\string\citation{apsrev41Control}}%
\fi
}%

\usepackage{graphicx}
\usepackage{varioref}
\usepackage{xr-hyper}
\usepackage{xcolor}
\usepackage{hyperref}
\hypersetup{colorlinks,linkcolor=blue,urlcolor=blue,citecolor=blue}

\usepackage{ulem}
\usepackage{braket}
\usepackage{array}
\usepackage{float}
\usepackage{multirow}
\newcolumntype{M}[1]{>{\centering\arraybackslash}m{#1\textwidth}}

\usepackage{subfiles}

\begin{document}

\title{Weak-signal extraction enabled by deep-neural-network denoising of diffraction data}

\author{J. Oppliger}
\affiliation{Physik-Institut, Universit\"{a}t Z\"{u}rich, Winterthurerstrasse 190, CH-8057 Z\"{u}rich, Switzerland}

\author{M. M. Denner}
\affiliation{Physik-Institut, Universit\"{a}t Z\"{u}rich, Winterthurerstrasse 190, CH-8057 Z\"{u}rich, Switzerland}

\author{J. K{\"u}spert}
\affiliation{Physik-Institut, Universit\"{a}t Z\"{u}rich, Winterthurerstrasse 190, CH-8057 Z\"{u}rich, Switzerland}

\author{R. Frison}
\affiliation{Physik-Institut, Universit\"{a}t Z\"{u}rich, Winterthurerstrasse 190, CH-8057 Z\"{u}rich, Switzerland}

\author{Q. Wang}
\affiliation{Physik-Institut, Universit\"{a}t Z\"{u}rich, Winterthurerstrasse 190, CH-8057 Z\"{u}rich, Switzerland}
\affiliation{Department of Physics, The Chinese University of Hong Kong, Shatin, Hong Kong, China}

\author{A. Morawietz}
\affiliation{Physik-Institut, Universit\"{a}t Z\"{u}rich, Winterthurerstrasse 
190, CH-8057 Z\"{u}rich, Switzerland}

\author{O. Ivashko}
%\email{oleh.ivashko@desy.de}
\affiliation{Deutsches Elektronen-Synchrotron DESY, Notkestra{\ss}e 85, 22607 Hamburg, Germany.}

\author{A.-C. Dippel}
%\email{oleh.ivashko@desy.de}
\affiliation{Deutsches Elektronen-Synchrotron DESY, Notkestra{\ss}e 85, 22607 Hamburg, Germany.}

\author{M.~v.~Zimmermann}
\affiliation{Deutsches Elektronen-Synchrotron DESY, Notkestra{\ss}e 85, 22607 Hamburg, Germany.}

\author{I. Biało}
\affiliation{Physik-Institut, Universit\"{a}t Z\"{u}rich, Winterthurerstrasse 190, CH-8057 Z\"{u}rich, Switzerland}
\affiliation{AGH University of Science and Technology, Faculty of Physics and Applied Computer Science, 30-059 Krak\'{o}w, Poland}

\author{L. Martinelli}
\affiliation{Physik-Institut, Universit\"{a}t Z\"{u}rich, Winterthurerstrasse 190, CH-8057 Z\"{u}rich, Switzerland}

\author{B. Fauqué}
\affiliation{JEIP, USR 3573 CNRS, Collège de France, PSL University,
11, place Marcelin Berthelot, 75231 Paris Cedex 05, France}

\author{J. Choi}
\affiliation{Diamond Light Source, Harwell Campus, Didcot, Oxfordshire OX11 0DE, United Kingdom}

\author{M.~Garcia-Fernandez}
\affiliation{Diamond Light Source, Harwell Campus, Didcot, Oxfordshire OX11 0DE, United Kingdom}

\author{K.-J.~Zhou}
\affiliation{Diamond Light Source, Harwell Campus, Didcot, Oxfordshire OX11 0DE, United Kingdom}

\author{N.~B.~Christensen}
\affiliation{Department of Physics, Technical University of Denmark, DK-2800 Kongens Lyngby, Denmark}

\author{T.~Kurosawa}
\affiliation{Department of Physics, Hokkaido University - Sapporo 060-0810, 
Japan}
 
\author{N.~Momono}
\affiliation{Department of Physics, Hokkaido University - Sapporo 060-0810, 
Japan}
\affiliation{Department of Applied Sciences, Muroran Institute of Technology, Muroran 050-8585, Japan}

\author{M.~Oda}
\affiliation{Department of Physics, Hokkaido University - Sapporo 060-0810, 
Japan}

\author{F.~D.~Natterer}
\affiliation{Physik-Institut, Universit\"{a}t Z\"{u}rich, Winterthurerstrasse 190, CH-8057 Z\"{u}rich, Switzerland}

\author{M.~H.~Fischer}
\affiliation{Physik-Institut, Universit\"{a}t Z\"{u}rich, Winterthurerstrasse 190, CH-8057 Z\"{u}rich, Switzerland}

\author{T. Neupert}
\affiliation{Physik-Institut, Universit\"{a}t Z\"{u}rich, Winterthurerstrasse 190, CH-8057 Z\"{u}rich, Switzerland}

\author{J.~Chang}
\affiliation{Physik-Institut, Universit\"{a}t Z\"{u}rich, Winterthurerstrasse 190, CH-8057 Z\"{u}rich, Switzerland}

\begin{abstract}
\textbf{
Removal or cancellation of noise has wide-spread applications for imaging and acoustics. In every-day-life applications, 
denoising may even include generative aspects, which are unfaithful to the ground truth. For scientific %\sout{applications} 
use, however, denoising must reproduce the ground truth accurately. Here, we show how data can be denoised via a deep convolutional neural network such that weak signals appear with quantitative accuracy.
In particular, we study 
X-ray diffraction on crystalline materials. 
We demonstrate that weak signals stemming from charge ordering, insignificant in the noisy data, become visible and accurate in the denoised data. This success is enabled by supervised training of a deep neural network with pairs of \emph{measured} low- and high-noise data. 
%This way, the neural network learns about the statistical properties of %\sout{the} 
%\cyan{actual} noise. 
We demonstrate that using artificial noise 
does not yield such quantitatively accurate results. Our approach thus illustrates a practical strategy for noise filtering that can be applied to challenging acquisition problems.}
\end{abstract}

\maketitle

In recent years, significant progress through deep learning techniques has been made in the field of image restoration~\cite{Jain2009,Zhang2017,Zhang2018,Noise2Noise1,Lefkimmiatis2018,Tian2020}. A central task in image restoration is %\cyan{pixel noise removal}
removing noise from an image~\cite{KimRSI2021,Schuetzke2022,zhang_poisson-gaussian_2019,STMnoise,elad_image_2023}, where pixel $j$ is composed of the intrinsic signal $s_j$ and noise $n_j$, $x_j=s_j+n_j$. A typical benchmark problem has correlated signal between neighboring pixels whereas the noise is uncorrelated and white.
Such denoising problems have been the subject of both supervised~\cite{Zhang2017,Zhang2018} and unsupervised~\cite{Noise2Noise1,krull_noise2void_2019,Batson2019} machine-learning approaches. Supervised algorithms rely on 
either ground-truth %\sout{pairs} 
$(x_j,s_j)$ or noise-2-noise training pairs $(x_j,x_j')$.
%\sout{training pairs and come in two flavors: either ground-truth pairs $(x_j,s_j)$ or a noise-2-noise approach is used.}
In the latter case, the image pairs %\sout{$(x_j,x_j')$} 
have different (or equal) noise levels. % \sout{of noise}. 
In both cases, deep convolutional neural networks (CNNs) have been successfully applied to images with Gaussian noise~\cite{Jain2009,Zhang2017,Zhang2018,Noise2Noise1,Lefkimmiatis2018}. Unsupervised approaches, sometimes dubbed noise-2-self, noise-2-void~\cite{krull_noise2void_2019}, or noise-as-clean~\cite{Xu2020}, have also been employed.
%\sout{Such approaches rely} c
Their realization relies on less training information %\sout{in the training data set} 
as a ground truth is absent. %\sout{Therefore, such} 
Unsupervised approaches %\sout{typically} 
therefore deliver (slightly) inferior performance compared to supervised algorithms.

Many scientific disciplines utilize 
digital data recording. %\sout{Whether a data structure is} 
One, two, or three dimensional data structures, %\sout{it} 
can always be transformed into a pixel-based picture format. Two-dimensional detectors are common across experimental fields such as astronomy, material science and medical imaging. 
Counting of events with time-independent probability are expected to follow Poisson statistics. As such, virus-cell infection, radioactivity and particle scattering are events following a Poisson distribution. That is, the signal $s_j$ and the noise $n_j$ are no longer independent %\sout{because} 
as the standard deviation %\sout{of a signal $s_j$} 
is given by $\sigma_j=\sqrt{s_j}$. Poisson noise can generally be %\cyan{thus} 
reduced by sufficient acquisition time. 
However, long exposure times are not always possible. %\sout{or desired}. 
For example: for radiation of molecules, proteins, or human tissue, low exposure times are required to avoid beam damage~\cite{Leijten2017}. %\sout{In other fields, experiments are carried out under short time scales.} 
Diffraction experiments in pulsed magnetic fields, %\sout{for example} 
by construction, have limited counting times and hence suffer from low-counting statistics~\cite{Knafo2016,RuffPRL2012}. Finally, many experiments explore multi-dimensional parameter spaces that are virtually impossible to cover completely with sufficient statistics. Thus there is a clear potential in developing robust methods to denoise low-counting (LC) statistics data %\sout{in a manner, which} 
to produce %\sout{s} 
results of comparable quality to what would be obtained from high-counting (HC) statistics data. %\sout{In that sense} 
By extension, noise filtering can speed-up exploratory approaches by orders of magnitudes.

\begin{figure*}
\centering
\includegraphics[width=0.99\textwidth]{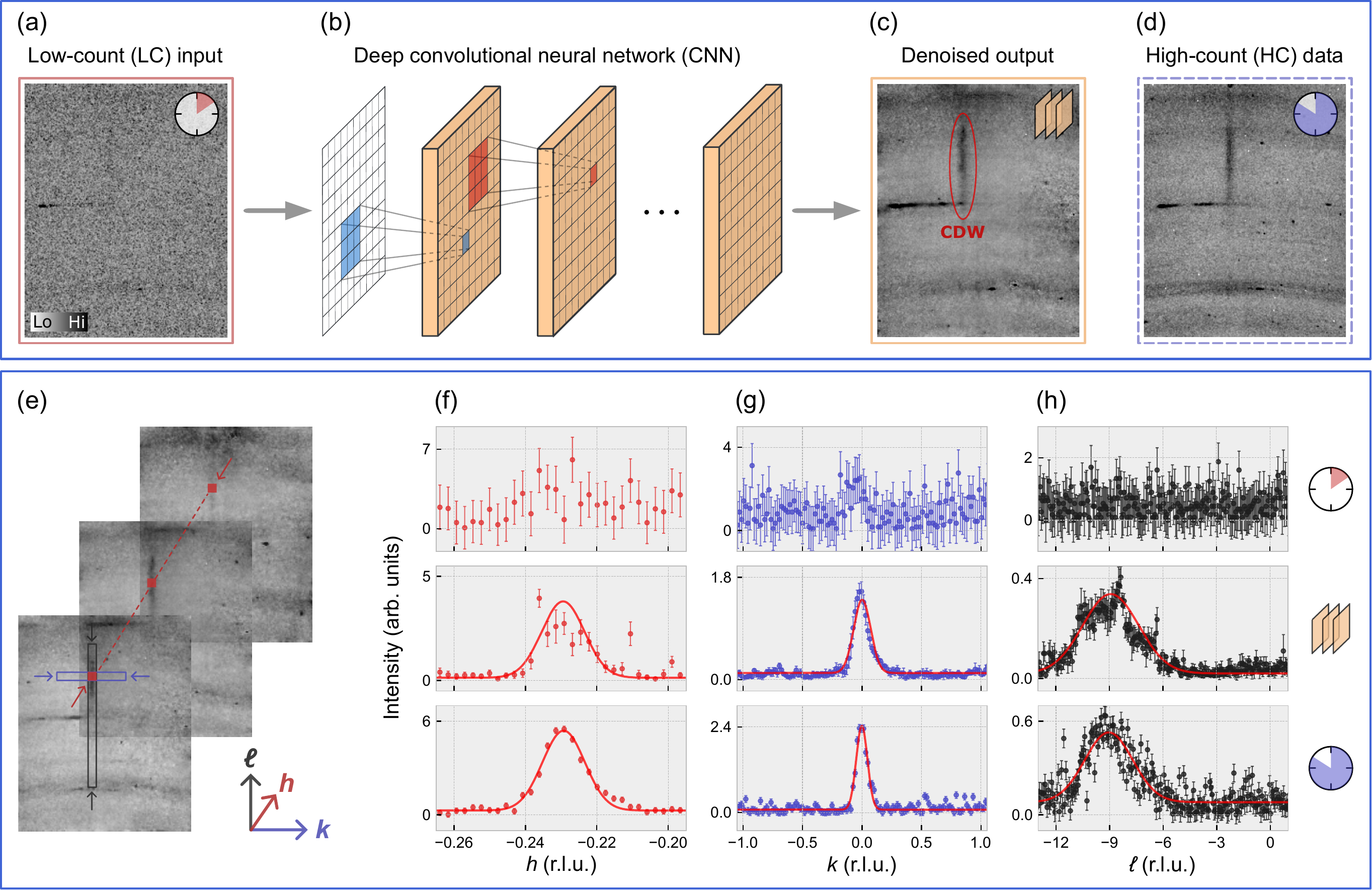}
\caption{Example of denoising X-ray diffraction (XRD) data using a deep convolutional neural network (CNN).
(a) Real experimental low-count  frame (exposure time 1 second) is used as an input to a deep CNN (b) -- trained to remove the noise. (c) The denoised output reveals a charge-density-wave signal (CDW, marked in red), barely visible in the raw low-count data. (d) The real experimental high-count frame (exposure time 20 seconds) is shown for comparison. (e) A stack of denoised X-ray intensity frames as in (c). Arrows indicate the projected reciprocal coordinates $Q=(h,k,\ell)$. (f-h) One-dimensional projected scans through  $Q \approx (0.23,0,8.5)$ along the $h$, $k$ and $\ell$ reciprocal space axes, in units of reciprocal lattice units (r.l.u.). For every projected scan, %\sout{a background close to the region of interest has been subtracted} 
a background subtraction has been performed -- see main text. Gaussian fits for high-count and denoised output profiles are indicated by red solid lines. The data points depicted in the denoised output profile are computed as the mean value over five training runs of the IRUNet neural network with different initial conditions. 
Error bars for low- and  high-count are shown under the assumption of counting statistics. Error bars for the denoised output are shown as the standard deviation over the mentioned training runs. %five training runs of the IRUNet neural network with different initial conditions.
The clock symbols indicate relative counting time and the network symbol indicates the denoised low-count produced by the neural network.}
\label{fig:denoising_schematic}
\end{figure*}

%\sout{Experimental data is challenging to} %\sout{noise filter} 
%\sout{denoise due to multiple noise sources such as, for example, read-out noise and counting statistics.}
However, removal of noise from experimental data is challenging. This can be attributed to the fact that
%\sout{As such,} 
experimental noise is the sum of %\sout{multiple noise statistics.}
%\sout{Noise filtering of %\sout{actual} 
%experimental data is challenging %\sout{because} 
%\sout{\cyan{due to the presence of}} %\sout{virtually all experimental data include} 
multiple noise sources, such as Poisson and read-out noise, %\sout{yielding} 
with %\sout{noise with} 
their respective statistical properties. It is therefore difficult -- often impossible -- to accurately simulate experimental noise. The common approach to analyze artificially added noise $n_j$ to a ground-truth signal $s_j$ is not directly applicable. Limited experimental training data however prevents further progress on this important problem~\cite{Bansal2022,WangScarceData2021}.

Here, we present experimental training data recorded by X-ray diffraction (XRD), $(x_j,x_j') = (x_j^{\text{LC}},x_j^{\text{HC}})$, where LC and HC refer to low- and high-counting statistics, respectively. 
Two %\sout{different} 
deep CNNs are trained %\cyan{in a supervised-fashion} 
on such pairs to remove noise from the LC data. %\sout{in a supervised way}. 
In a further step, the performance of the neural networks trained on %\sout{actual} 
experimental data is compared to the same networks trained on artificial training pairs where the HC data is corrupted with synthetic Poisson noise. 
We find that noise filtering of experimental noise -- the ultimately %\sout{important} 
relevant task -- is significantly improved by training the neural networks on experimental data. This fact is particularly evident when analyzing physical length scales  associated with weak signals. As such, we provide a noise filtering approach for scientific data with challenging signal-to-noise features. 

\begin{figure*}
\centering
\includegraphics[width=\textwidth]{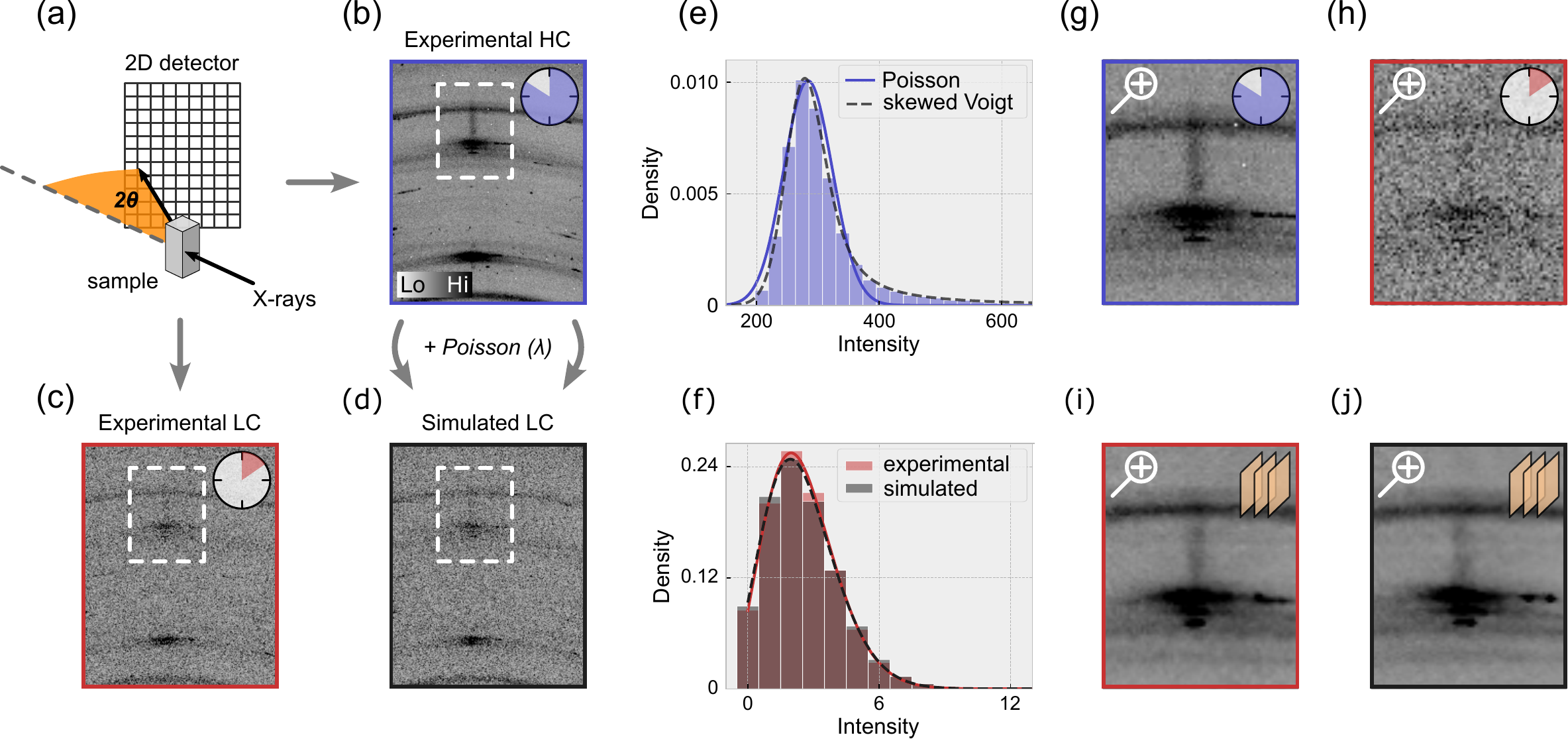}
\caption{Comparison of %\sout{real} 
experimental and simulated noise statistics. (a)  Schematic of the  experimental X-ray diffraction setup. Long exposure time leads to a high-count (HC) frame (b) while short exposure time leads to a low-count (LC) frame (c). Adding Poisson noise to experimental high-count frame (b) leads to a simulated low-count frame (d). (e) Intensity distribution of the high-count frame (b) with fitted Poisson and skewed Voigt profiles. (f) Intensity distribution of the experimental and simulated low-count frame in (c,d) with fitted Poisson profile.  (g,h) Zoom of white dashed rectangular region in (b,c) encircling the charge-density-wave reflection. (i,j) Zoom of white dashed rectangular region in (c,d) after denoising using the IRUNet network trained on the respective noise distributions.}
\label{fig:noise_statistic_comparison}
\end{figure*}

\section{Result section}

\subsection{X-ray diffraction data}

An example of %\sout{X-ray diffraction} 
XRD intensities recorded on the high-temperature superconductor La$_{1.88}$Sr$_{0.12}$CuO$_4$
is shown in \mbox{Figure~\ref{fig:denoising_schematic}(a,d)} and \mbox{\ref{fig:noise_statistic_comparison}(b,c)} with LC and HC frames, respectively. The experimental setup is schematically depicted in \mbox{Figure~\ref{fig:noise_statistic_comparison}(a)} and further described in %\sout{and a detailed description is given in} \cyan{-- see also} 
the  methods section. Although the data %\sout{recording} 
covers volumes of reciprocal space, the training is carried out on two-dimensional slices (so-called frames). %\purple{The CNN kernels are therefore two-dimensional  correlation pixels across different frames.}
Therefore, the neural networks do not have access to the three-dimensionality of the data set but rely on the two-dimensional correlation of pixels in individual frames. The LC (HC) data are recorded typically for 1 (20)~seconds.
Such experimental pairs were recorded successively with all experimental parameters fixed. %\sout{The complete} \sout{Our} 
The entire data set contains 7134 frame pairs %\sout{of} 
(194 x 242 pixels each) %The data \sout{is composed of} 
%\sout{It} 
%\purple{and} 
and includes signals with intensities varying over six orders of magnitude. Weak two-dimensional charge-density-wave (CDW) order~\cite{Christensen14,CroftPRB2014,ThampyPRB2014} %\sout{are reflected by correspondingly low intensities forming} 
manifests by %\sout{a} 
vertical rod-like shapes. In cuprates, the exact nature of CDW ordering is still being debated. On an atomistic level, the CDW in La$_{1.88}$Sr$_{0.12}$CuO$_4$ represents monoclinic distortions of the fundamental orthorhombic crystal structure~\cite{frison_crystal_2022}. Fundamental Bragg peaks (not shown) are more intense and distributed circularly over much fewer pixels. The data also contains Debye-Scherrer (powder) rings originating from the polycrystalline %\sout{components of the} 
sample environment. Finally, the data includes spurions (unidentified signal) and dead pixels. %\sout{Signals that originate from} 
Bragg scattering %\sout{would change} 
implies a direct connection between scattering angle (i.e. %\sout{move} 
position on the detector) and incident photon energy/wavelength. %\sout{upon change of energy/wavelength of the incident photons}. 
As such, conclusions drawn here are invariant under different scattering angles defined by incident photon energy or sample (lattice parameter).
%As such, conclusions drawn here are invariant under different scattering angles defined by \sout{choice of sample or} \cyan{the} incident photon energy.
%\sout{This invariance also extends to}
Furthermore, a different doping concentration in our La$_{2-x}$Sr$_{x}$CuO$_4$ show case,   %could 
%\sout{modifies} 
would change the charge order incommensurability~\cite{HuckerPRB2014}, but %\sout{will} 
not %\sout{change} 
the overall data content. %\sout{The complete} 

The data set is separated into a training, validation, and test set. All frames containing obvious CDW signals -- our main feature-of-interest -- are excluded from the training and validation set. %\sout{All} 
These frames are instead moved to %\sout{a separate} 
the test set, which is
%, \sout{\cyan{which \sout{will be} \purple{is}}} 
used for performance evaluation. Overall, the size of the training set is 3280 pairs while the size of the validation set is 820 pairs. 

%\sout{\purple{Once trained, the network can be used to noise-filter}}
%\sout{\purple{future collected data. In this work, we demonstrate}}
%\sout{\purple{the network stability on resonant inelastic x-ray data.}}

\begin{figure*} 
\centering
\includegraphics[width=0.95\textwidth]{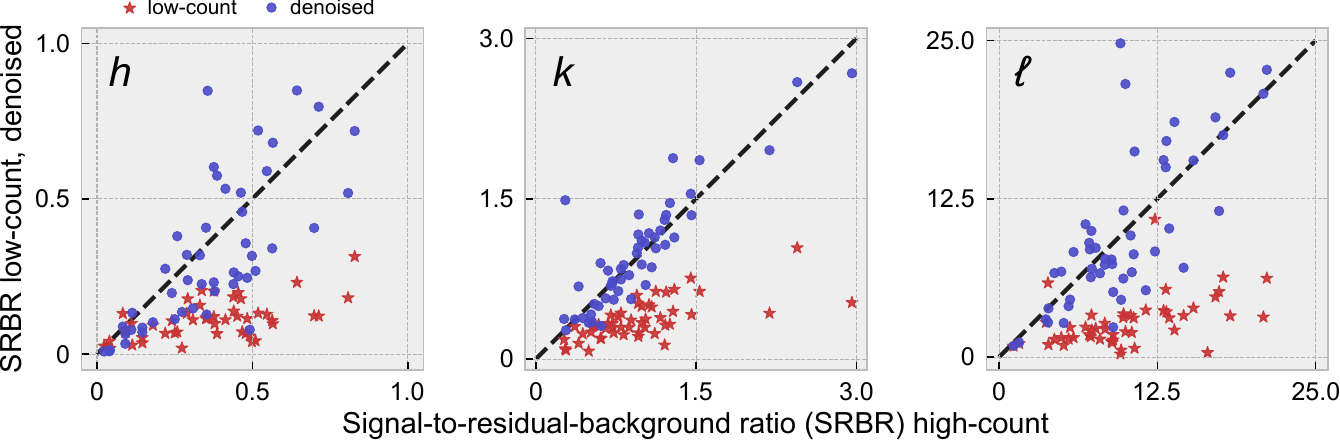}
\caption{Enhancement of signal-to-residual-background ratio (SRBR) using CNN denoising via the IRUNet network trained on experimental data. Multiple frames containing charge-density-wave signals are analyzed along the reciprocal $(h,k,\ell)$ direction in a similar fashion as in \mbox{Figure~\ref{fig:denoising_schematic}(f-h)}. The signal-to-residual-background ratio of the charge-density-wave reflection in the high-count frame is plotted against the signal-to-residual-background ratio of the low-count frame and its denoised version.  We observe that the denoising of the low-count frames improves the signal-to-residual-background ratio and, in many cases, even leads to better results than the high-count data.}
\label{fig:performance_comparison}
\end{figure*}

\subsection{Artificial noise generation}
As shown in \mbox{Figure~\ref{fig:noise_statistic_comparison}}(f), the experimental LC data %\sout{is following} 
follows an approximate Poisson distribution. Therefore, to complement the experimental LC data, we artificially create LC data by adding Poisson noise to the experimental HC data. %\sout{X-ray diffraction} 
XRD data is governed by counting statistics, where the probability of a single photon hitting pixel $j$ is theoretically given by the Poisson probability distribution $P_j$ %\sout{$P_{i,k}$} 
for large total count $N$ -- ideally $N\rightarrow \infty$. For fair comparison, artificial and experimental LC data should be statistically similar. To achieve this, we define $\lambda_f$ as the ratio between frame-integrated low-count $N_{f}^{\text{LC}}$ and high-count $N_{f}^{\text{HC}}$ ($f$ here refers to a frame index) and $\lambda=\text{median}(\lambda_f)$. %\sout{The median value of $\lambda_f$ over all pairs of experimental training data is labeled $\lambda$.} \sout{We normalize} 
Each HC frame %\sout{by multiplication} 
is then normalized with $\lambda$ and %\sout{then generate a} 
LC frames being generated by adding the associated Poisson noise, resulting in simulated LC frames as shown in \mbox{Figure~\ref{fig:noise_statistic_comparison}(d)}. 
%\sout{In \mbox{Figure~\ref{fig:noise_statistic_comparison}(c,d,f)}, we show that the experimental LC data %\sout{is following} 
%\cyan{follows} an approximate Poisson distribution. }
Notice that signal intensities may vary by many orders of magnitude across the detector pixels, therefore the HC data typically displays an asymmetric probability distribution as shown in \mbox{Figure~\ref{fig:noise_statistic_comparison}(e)}. 
 
\subsection{Deep neural-network architectures and training}
We implement two %\sout{different} 
neural-network architectures referred to as VDSR~\cite{kim_accurate_2016} and IRUNet~\cite{gil_zuluaga_blind_2021} -- see \mbox{Figure~\ref{fig:denoising_schematic}(a-c)} for a schematic illustration. 
%\sout{A schematic illustration of the working principle of the denoising networks is shown in \mbox{Figure~\ref{fig:denoising_schematic}(a-c)}.} 
%\sout{LC frames serve as an input to a deep CNN.} 
The networks %\sout{then} 
learn the intrinsic features of the LC input frames, and produce a denoised output using the HC frames as reference. The VDSR architecture relies on stacking many convolutional layers and uses a residual learning approach to extract the noise-free data from its noisy variant~\cite{Zhang2017,he_deep_2016}. The IRUNet architecture combines convolutional layers with an encoder/decoder framework, utilizing %\sout{different types of} 
skip connections to reduce the vanishing gradient problem and increase accuracy. An Adam optimizer~\cite{kingma_adam_2017} with the AMSGrad variant~\cite{reddi_convergence_2019} is used %\sout{for training the networks and} 
to improve convergence. %\sout{During the data loading  phase both LC and HC} 
All frames are normalized by their total intensity, %\sout{This} \sout{ensures}
ensuring %\sout{an} 
equal scaling between LC and HC frames. During training, we apply data augmentation in the form of mirroring the frames along the $\ell$ and $k$ direction, %\sout{rotation by $\pm \, \text{90 degrees}$} 
and randomly adjusting the global brightness of the frames. Additional information can be found in the methods section. 
%\sout{A comprehensive review on image denoising and noise-filtering networks is given in Ref.~\cite{elad_image_2023}.}
%\sout{As mentioned previously, we exclude data pairs containing clearly visible CDW order reflections in the training and validation set. As such the data with CDW reflections is used for performance evaluation. Additional information on the networks as well as training and performance measures are provided in the methods section.}

\begin{table*}
\renewcommand{\arraystretch}{1.5}
\centering
\caption{Average Gaussian fitting results of different training and evaluation protocols using multiple frames from the test set containing charge-density-wave (CDW) signals. The first column indicates the used training and evaluation methodology. For example training on artificial Poisson noise and evaluation on experimental noise (Poisson $\rightarrow$ Exp.). The Gaussian peak position $\mu_\alpha$ and standard deviation $\sigma_\alpha$ with $\alpha = (h,k,\ell)$ are given as the mean absolute error between the Gaussian parameter obtained from the denoised and the one obtained from the high-count signal (the lower the better). Values for the signal-to-residual-background ratio (SRBR$_\alpha$) are given as the absolute ratio of the Gaussian parameter obtained from the denoised and the one obtained from the high-count signal (the higher the better). Values for $\mu_\alpha$ as well as values for $\sigma_\alpha$ are scaled as indicated. Because of the broader peak in the $\ell$ direction, a scaling of 10 for $\mu_\ell$ and $\sigma_\ell$ has been chosen over a scaling of 100.}
\label{tab:performance_table}

\vspace{0.5cm}

\begin{tabular}{M{0.15}|M{0.085}|M{0.085}|M{0.085}|M{0.085}|M{0.085}|M{0.085}|M{0.085}|M{0.085}|M{0.085}|}
\cline{2-10}
 & $\mu_h$ ($\times 10^2$) & $\mu_k$ ($\times 10^2$) & $\mu_\ell$ ($\times 10$) & $\sigma_h$ ($\times 10^2$) & $\sigma_k$ ($\times 10^2$) & $\sigma_\ell$ ($\times 10$) & SRBR$_h$ & SRBR$_k$ & SRBR$_\ell$\\
\hline
\multicolumn{1}{|c|}{Low-count} & 0.66 (05) & 1.94 (13) & 2.48 (37) & 0.18 (04) & 1.17 (13) & 2.85 (54) & 0.39 (07) & 0.41 (04) & 0.31 (07)\\
\hline
\end{tabular}

\vspace*{0.25cm}

\begin{tabular}{M{0.15}|M{0.085}|M{0.085}|M{0.085}|M{0.085}|M{0.085}|M{0.085}|M{0.085}|M{0.085}|M{0.085}|}
\multicolumn{10}{c}{IRUNet}\\
\hline
\multicolumn{1}{|c|}{Poisson $\rightarrow$ Poisson} & 0.29 (03) & 0.56 (09) & 1.08 (12) & 0.14 (02) & 0.75 (10) & 1.03 (15) & 0.90 (10) & 0.96 (03) & 1.13 (04)\\
\multicolumn{1}{|c|}{Poisson $\rightarrow$ Exp.} & 0.75 (04) & 0.87 (11) & 4.17 (14) & 0.31 (04) & 1.65 (13) & 1.37 (17) & 1.02 (25) & 0.95 (13) & 0.97 (04)\\
\multicolumn{1}{|c|}{Exp. $\rightarrow$ Exp.} & \textbf{0.19 (03)} & \textbf{0.65 (07)} & \textbf{1.47 (12)} & \textbf{0.31 (03)} & \textbf{0.69 (08)} & \textbf{1.50 (15)} & \textbf{1.41 (24)} & \textbf{1.00 (02)} & \textbf{1.21 (05)}\\
\hline
\end{tabular}

\vspace*{0.25cm}

\begin{tabular}{M{0.15}|M{0.085}|M{0.085}|M{0.085}|M{0.085}|M{0.085}|M{0.085}|M{0.085}|M{0.085}|M{0.085}|}
\multicolumn{10}{c}{VDSR}\\
\hline
\multicolumn{1}{|c|}{Poisson $\rightarrow$ Poisson} &  0.38 (02) & 0.58 (08) & 1.01 (11) & 0.14 (02) & 0.78 (09) & 1.23 (14) & 0.95 (08) & 1.05 (02) & 1.22 (05)\\
\multicolumn{1}{|c|}{Poisson $\rightarrow$ Exp.} & 0.46 (03) & 0.72 (14) & 2.62 (11) & 0.19 (02) & 1.28 (18) & 1.18 (14) & 0.94 (07) & 1.15 (03) & 1.24 (05)\\
\multicolumn{1}{|c|}{Exp. $\rightarrow$ Exp.} & \textbf{0.32 (02)} & \textbf{0.63 (08)} & \textbf{1.11 (11)} & \textbf{0.16 (02)} & \textbf{0.73 (08)} & \textbf{0.95 (14)} & \textbf{0.97 (09)} & \textbf{1.09 (02)} & \textbf{1.25 (04)}\\
\hline
\end{tabular}
\end{table*}

\subsection{Analysis}

%\sout{In \mbox{Figure~\ref{fig:noise_statistic_comparison}(c,d,f)}, we show that the experimental LC data %\sout{is following} 
%\cyan{follows} an approximate Poisson distribution. Conversely, as signal\sout{s} \cyan{intensities} may vary by many orders of magnitude across the detector pixels, the HC data typically displays an  asymmetric probability distribution as shown in \mbox{Figure~\ref{fig:noise_statistic_comparison}(e)}.} 
Performance evaluation of the trained neural networks -- on test data -- is illustrated by
%\sout{To evaluate the denoising performance of the trained neural networks,
%we %\sout{are studying} 
%construct} 
one-dimensional line-cuts (along the reciprocal $h$, $k$, and $\ell$ directions) through the CDW ordering vector -- see \mbox{Figure~\ref{fig:denoising_schematic}(e-h)}. This involves the summation of pixel intensities within a region-of-interest (ROI) and subtraction of neighbouring ROIs -- see supplementary information. This subtraction is made to eliminate the background surrounding the CDW signal such as powder rings. As the background subtraction is not always perfect, the one-dimensional line-cuts are %\sout{therefore} 
composed of signal and a small residual "background". To avoid negative residual background intensities, a small constant shift has been applied. %Since the network can cope with negative intensities, we have added small shift.}
%\sout{to  reduce the influence of powder rings.}
%\sout{The denoised output -- including the CDW signal along the three reciprocal space directions -- illustrates the benefit of the denoising network applied to the LC data. The results shown in \mbox{Figure~\ref{fig:denoising_schematic}} are obtained upon applying the trained neural network on previously unseen data from the %\sout{separate} 
%test set.}
%\sout{As Shown} 
In \mbox{Figure~\ref{fig:denoising_schematic}(f-h)}, we  analyze the line-cuts %\sout{signal properties} 
by fitting a Gaussian model.
%\sout{with a constant \cyan{residual} background to the projected one-dimensional line-cuts}. 
The resulting parameters are the amplitude $A$, the peak position $\mu$, the standard deviation $\sigma$, and the constant residual background $C$. We furthermore define the signal-to-residual-background ratio \mbox{SRBR $= A/C$}. In \mbox{Figure~\ref{fig:performance_comparison}}, we show the SRBR for 50 different examples of CDW order from the %\sout{previously mentioned} 
test set. Denoising using a neural network significantly improves the SRBR of the CDW order -- oftentimes surpassing the results obtained from the HC data. Due to the random nature of noise, the network is not able to learn the small but finite noise component of the HC data, resulting in an efficient noise-removal. These results are summarized in \mbox{Table~\ref{tab:performance_table}} and put in contrast to values extracted from the unfiltered LC data. We also compare the training with experimental and artificial noise %\sout{with} 
of similar noise levels. Additionally to the SRBR, we calculate the mean absolute error between the denoised peak position $\mu_{h,k,\ell}$ and standard deviation $\sigma_{h,k,\ell}$ with the HC values. From these results, we %\sout{can} 
conclude that training on experimental data greatly %\sout{enhances the quality of the} 
improves noise filtering. This conclusion holds even in the case when the amount of artificial training data is larger than the amount of experimental training data. A considerable improvement can be achieved by employing a multiscale training procedure where the artificial training data covers a wide range of statistics. More details as well as a table containing standard image quality metrics describing the denoising performance can be found in the supplementary information. Finally, we observe that both IRUNet and VDSR networks, on average, achieve comparable results despite their different architectures.

\section{Discussion}
%\sout{From} 
Eventually, removal of experimental noise is the goal of noise-filtering algorithms. 
Many studies have focused on filtering artificial noise from photographs~\cite{Jain2009,Zhang2017,Zhang2018,Noise2Noise1,Lefkimmiatis2018,Tian2020}. The artificial noise typically has a single statistical distribution (Gaussian, Poisson, or Bernoulli) and the photographs are represented by Red-Green-Blue (RGB) color scales from 0 to 255. Experimental %\sout{data} 
noise poses a much harder problem as multiple noise sources are present and the signal can vary by many orders of magnitude. This suggests that denoising algorithms should (also) be benchmarked on more challenging experimental data. Here, we provide an %\sout{X-ray diffraction} 
XRD %\sout{training} 
data set where the signal intensity varies by six orders of magnitude. Not surprisingly, we find that networks trained to remove artificial noise perform well on exactly that task. However, the high performance, unfortunately, does not carry over to the filtering of experimental noise. Our results suggest %\sout{we conclude} 
that neural networks filter experimental noise better when trained on experimental noise rather than artificial noise profiles. This statement is especially true when the noise levels of experimental and artificial noise %\sout{level is} 
are comparable.
%\sout{This statement is especially true when the noise level of experimental and artificial noise level is comparable. Noise filtering using CNNs strongly reduces the noise and oftentimes leads to an even better SBR as the actual high-count data. Additionally, we find that a CNN trained for denoising data with specific noise statistics can also enhance the signal quality of data with slightly different noise statistics.}

\begin{figure*} 
\centering
\includegraphics[width=\textwidth]{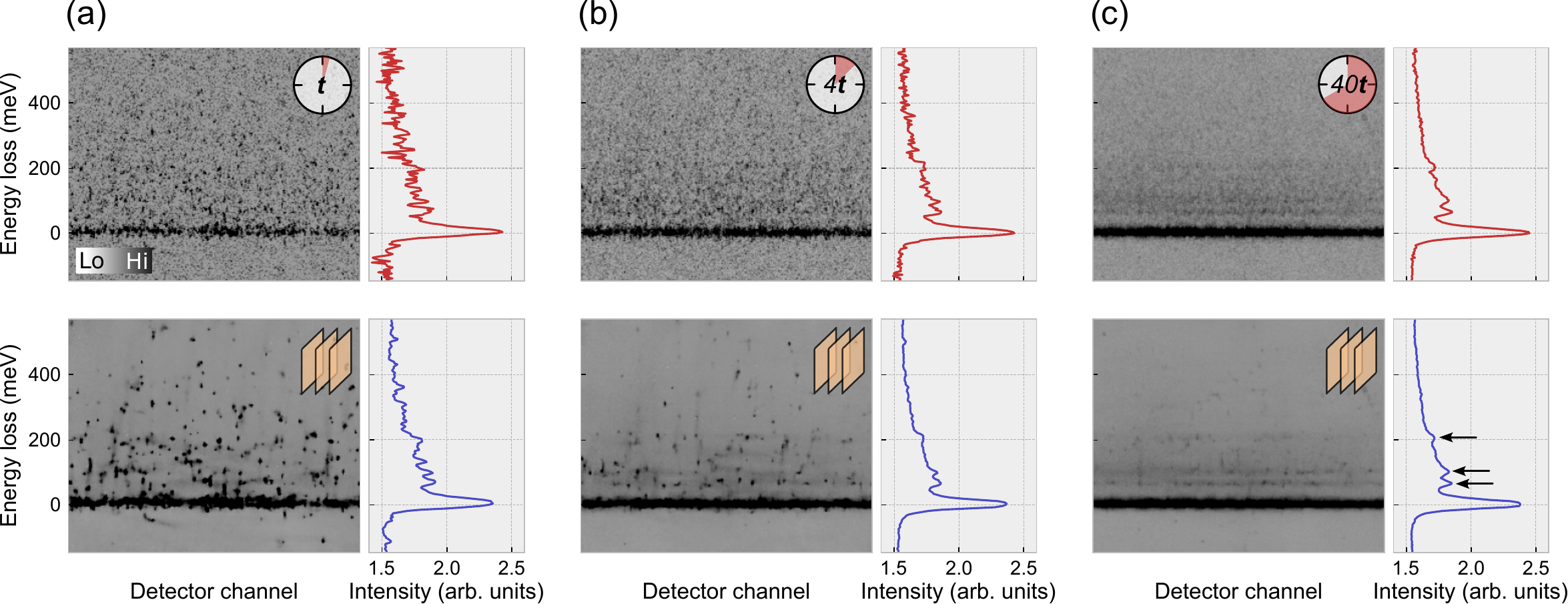}
\caption{Resonant inelastic X-ray scattering (RIXS) spectra recorded on SrTiO$_3$. (a-c) RIXS spectra with counting statistics of 1, 4 and 40 times 3 minutes (top row). Left panels display counting intensities %\purple{(false color scale) versus detector channel and energy loss}
with detector channel versus energy loss.
%\sout{on the two-dimensional detector whereas} 
The right panels show the (horizontally) projected RIXS spectra. The panels in the bottom row are the %\sout{respective} 
corresponding denoised neural-network outputs of the top row. Three inelastic peaks are highlighted by arrows in (c).}
\label{fig:rixs_denoising}
\end{figure*}

To illustrate the generality and robustness of this work, we apply %\sout{our} 
the trained network to resonant inelastic X-ray scattering (RIXS) data. %\sout{X-ray diffraction} 
XRD and RIXS are fundamentally different experimental techniques -- see methods section. %\sout{for additional information.} 
In \mbox{Figure~\ref{fig:rixs_denoising}}, we show a raw RIXS spectrum recorded on SrTiO$_3$ with different counting statistics as indicated (top panels). In the bottom panels we show the corresponding denoised output from a CNN  trained on experimental %\sout{X-ray diffraction} 
XRD data exclusively. %\sout{only}.
As the used RIXS detector does not offer single photon sensitivity, the signal is not expected to follow pure Poisson statistics. Despite the dissimilar experimental technique, different sample, and different noise distribution, the trained neural network achieves a visible noise reduction and consequently enhances the SRBR.
The successful denoising of RIXS data likely stems from the rich variation of signals (powder rings, charge order and lattice Bragg peaks, spurious, and dead pixels) and noise sources in the XRD training data.
%\sout{We expect that training the neural networks with additional RIXS data will lead to further performance enhancements at low-counting statistics.}

%Eventually, removal of experimental noise is the goal of noise filtering algorithms. 
%Many studies have focused on filtering artificial noise from photographs~\cite{Jain2009,Zhang2017,Zhang2018,Noise2Noise1,Lefkimmiatis2018,Tian2020}. The artificial noise typically has a single statistical distribution (Gaussian, Poisson, or Bernoulli) and the photographs are represented by Red-Green-Blue (RGB) color scales from 0 to 255. Experimental data poses a much harder problem as multiple noise sources are present and the signal can vary by many orders of magnitude. This suggests that denoising algorithms should (also) be benchmarked on more challenging experimental data. Here, we provide an \sout{X-ray diffraction} \cyan{XRD} \sout{training} data set where the signal intensity varies by six orders of magnitude. Not surprisingly, we find that networks trained to remove artificial noise perform well on exactly that task. However, the high performance, unfortunately, does not carry over to the filtering of experimental noise.

Our results, therefore, encourage the collection of even more diverse training data with different compositions of noise sources from other scattering, spectroscopy, and microscopy techniques. Small-angle neutron and (resonant inelastic) \mbox{X-ray scattering data~\cite{SANSreview}} would be an obvious %\sout{path}
choice to extend the training data. Spectroscopies~\cite{KimRSI2021,Schuetzke2022} and microscopies~\cite{zhang_poisson-gaussian_2019,STMnoise} such as angle-resolved photoemission electron spectroscopy~\cite{ARPES2021} and transmission electron microscopy~\cite{TEM2012,Mevenkamp2015} data could also help expanding the amount and variety of training data. Furthermore, the application of transfer learning~\cite{pan_survey_2010} using a pre-trained model might prove beneficial in reducing the amount of distinct training data needed. By applying our method to future studies, a large amount of beamtime could be saved, or a fixed beamtime budget could be used more efficiently by, for example, being able to probe a larger parameter space.

\section{Methods}
\textit{X-ray diffraction:} The training data was recorded on a La$_{1.88}$Sr$_{0.12}$CuO$_4$ single crystal~\cite{ChangPRB2008}, at beamline P21.1 at the PETRA III storage ring at DESY in Hamburg. The scattering intensities were recorded using a DECTRIS PILATUS3 X CdTe 100k detector. This detector provides us with 
195 x 487 pixels per frame and a bit-depth of
32. Each pixel is associated with a horizontal and a vertical scattering angle from which reciprocal space coordinates can be reconstructed as described in~\cite{frison_crystal_2022}. The convolutional neural network (CNN) training is independent of this reconstruction that is done to extract correlation lengths. The diffractometer was operated with 100~keV photons and the sample was cooled to around 30~K where a charge-density-wave (CDW) order is fully developed. The charge order has a short correlation length along the $c$-axis directions. Hence, along the reciprocal $c$-axis ($\ell$), the CDW order manifests by a long rod of scattering intensity.

\vspace*{0.1cm}

\textit{Resonant inelastic X-ray scattering:} The oxygen K-edge RIXS spectra  were recorded at the I21 beamline~\cite{Zhou2022} at Diamond Light Source on a SrTiO$_3$ crystal. Linear vertical light polarization and photon energy of $\sim$531~eV were used. The sample temperature was 20~K and the momentum transfer was set to \mbox{$(h,k,\ell) = (0, 0, 0.245)$ r.l.u.}.

\vspace*{0.1cm}

\textit{Convolutional neural networks:}
Although networks for three-dimensional data structures exist~\cite{cicek_3d_2016}, we employed architectures designed for uncorrelated two-dimensional images. A comprehensive review of deep learning and convolutional neural networks applied to noise-filtering of images is given in Ref.~\cite{elad_image_2023}. Generally, many networks are displaying comparable performance. In this work, we implemented two different neural-network architectures referred to as VDSR~\cite{kim_accurate_2016} and IRUNet~\cite{gil_zuluaga_blind_2021}. %\sout{were implemented}. 
For the VDSR architecture, we did not include the final addition layer as we do not find a significant performance change -- see supplementary information. %\sout{\mbox{Table~\ref{tab:hyperparameters_models}} lists additional training parameters used for the two models next to the ones described in the main text}. 
The weights of the convolutional layers are randomly initialized using the He method~\cite{he_delving_2015}. For the VDSR model we make use of a parametric rectifying linear unit (PReLU)~\cite{he_delving_2015} after each convolutional layer while a normal ReLU is used in the IRUNet architecture. The VDSR network has been trained for 150 epochs using a batch size of 8 and an initial learning rate of 5$\times$10$^{-4}$. The IRUNet network has been trained for 200 epochs using a batch size of 16 and an initial learning rate of 5$\times$10$^{-4}$. The learning rate was decreased after a certain number of epochs to ensure good convergence. For the VDSR model, we multiply the learning rate with 0.5 after every 50 epochs. For the IRUNet model, the learning rate is multiplied with 0.5 after 150 epochs. The total training duration of the VDSR and IRUNet models was on average around 20 and 10 hours respectively on a Nvidia Tesla P100 GPU with 10 GB of VRAM using TensorFlow~2.4.1. A discussion of the receptive field of the neural networks can be found in the supplementary information. %\purple{*Maybe move the entire discussion of the receptive field to the supplementary?*} 

\vspace*{0.1cm}

\textit{Loss function:} During each training epoch, the performance of the neural networks is determined by comparing the denoised output with the high-count frame. The used loss function $L$ is given by a combination of mean absolute error (MAE) and multiscale structural similarity (MSSIM)~\cite{zhao_loss_2017,KimRSI2021}. We find that this loss function results in a better overall denoising performance when compared to other losses such as mean squared error (L2 loss) -- see supplementary information. 

% \vspace*{-0.4cm}

% \begin{align*}
% L = (1 - \alpha) L_{\text{MAE}} + \alpha L_{\text{MS-SIM}} \quad \text{with} \quad \alpha = 0.7
% \end{align*}

% \noindent

%\purple{*maybe make extended figure instead of SI?*}.

% \vspace*{0.1cm}
\newpage

\textit{Data availability:}
The experimental data used in this work can be found at \href{https://doi.org/10.5281/zenodo.8237173}{doi.org/10.5281/zenodo.8237173}.

\vspace*{0.1cm}

\textit{Code availability:}
The code used for the setup and training of the neural networks is available at \href{https://github.com/joppli/XRD_denoising}{github.com/joppli/XRD\_denoising}.

\vspace*{0.1cm}

\textit{Competing interests:}
The authors declare no competing interests.

\vspace*{0.1cm}

\textit{Acknowledgements:} 
We thank Nicola Serra for GPU-time that has been used for training the neural networks, Alexander Steppke, Nik Dennler, and Younsik Kim for insightful discussions. We acknowledge DESY (Hamburg, Germany), a member of the Helmholtz Association HGF, for the provision of experimental facilities. Parts of this research were carried out at beamline P21.1 at PETRA III. J.O. acknowledges support from a Candoc grant of the University of Zurich (Grant no. K-72334-06-01). J.K., Q.W., and J.C. express gratitude to the Swiss National Science foundation for funding (SNF) under grant number 200021$\_$188564. Q.W. is supported by the Research Grants Council of Hong Kong (ECS No. 24306223). J.O. and F.D.N. thank the SNF (PP00P2$\_$176866) and ONR (N00014-20-1-2352) for generous support. I.B. and L.M. acknowledge support from the Swiss Government Excellence Scholarship under project numbers ESKAS-Nr: 2022.0001 and ESKAS-Nr: 2023.0052. M.M.D. and T.N. acknowledge support from the European Union’s Horizon 2020 research and innovation programme (ERC-StG-Neupert-757867-PARATOP). M.M.D. acknowledges support from a Forschungskredit of the University of Zurich (Grant no. FK-22-085). J.K., R.F., and A.M. were supported by the project
CALIPSOplus under the Grant Agreement 730872 from the EU Framework Programme for Research and Innovation HORIZON 2020. J.K. acknowledges funding from the German Academic Scholarship Foundation. N.B.C. thanks the Danish Agency for Science, Technology, and Innovation for funding the instrument center DanScatt and acknowledges support from the Q-MAT ESS Lighthouse initiative. Finally, we acknowledge Diamond Light Source for providing beamtime at I21 under Proposal MM31819.

\vspace*{0.1cm}

\textit{Author contributions:}
T.K., N.M., and M.O. grew and characterized the LSCO single crystals. The SrTiO$_3$ samples were obtained from commercial source. J.K., R.F., Q.W., A.M., N.B.C., O.I., A.-C.D. and M.v.Z carried out the XRD data collection. I.B., L.M., B.F., J.Choi, M.G-F. and K.Z. recorded the RIXS data. J.O. did the machine learning training and data analysis with supervision from M.M.D., F.D.N., M.H.F., T.N., and J.C.. All authors contributed to the manuscript text.

\bibliography{lsco_ref}

% \newpage
\clearpage

\section*{Supplementary information}

\subsection{Details about the loss function}

In this work we made use of a loss function that is known to perform well for images intended for evaluation by a human observer~\cite{zhao_loss_2017,KimRSI2021}. It combines a pixel-wise absolute error (mean absolute error, MAE) with local structural similarity that is calculated at different scales (multiscale structural similarity, MS-SSIM~\cite{wang_image_2004,wang_multiscale_2003})

\begin{figure*}
\centering
\includegraphics[width=0.9\textwidth]{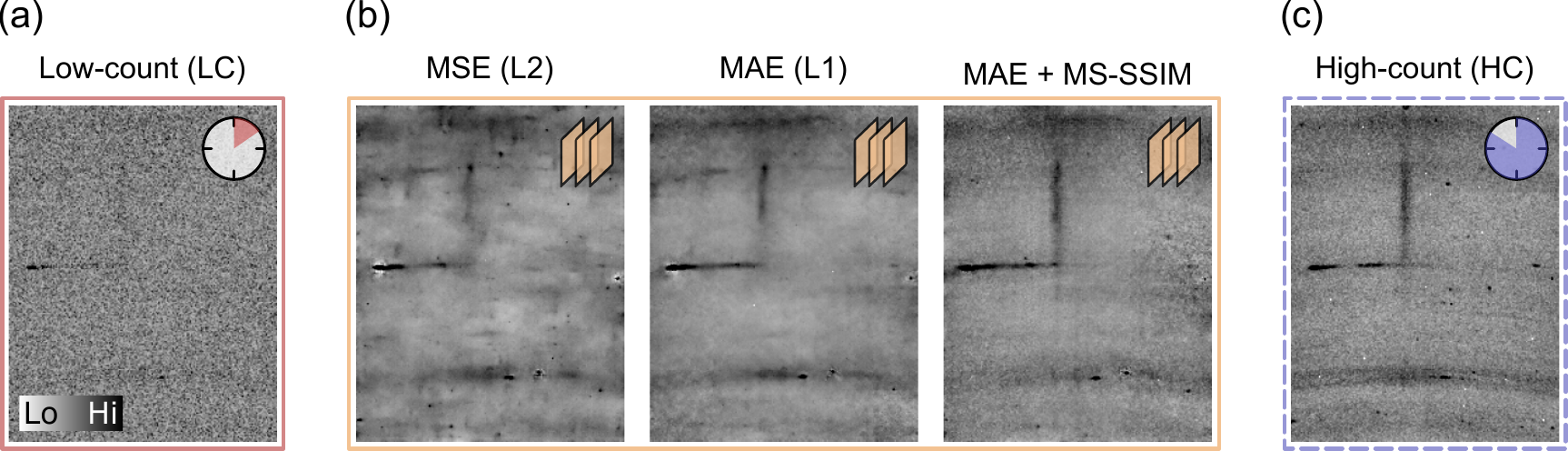}
\caption{Impact of different loss functions on the denoised neural-network output. (a) Low-count frame. (b) Denoised low-count frame in (a) using different loss functions during the training of the network. A combination of mean absolute error and multiscale structural similarity (MAE + MS-SSIM) shows the best denoising performance. (c) High-count frame for comparison.}
\label{fig:different_loss_functions}
\end{figure*}

\vspace{-0.4cm}

\begin{align*}
L = (1 - \alpha) L_{\text{MAE}} + \alpha L_{\text{MS-SIM}} \quad \text{with} \quad \alpha = 0.7
\end{align*}

where the value of $\alpha$ has been chosen empirically. In \mbox{Figure~\ref{fig:different_loss_functions}} we show the denoising performance on a single low-count (LC) frame when using aforementioned and other standard loss functions such as mean absolute error (MAE,~L1) and mean squared error (MSE,~L2). We observe that the MSE loss leads to poor performance, resulting in a locally smeared background with only minimal structural patterns. MAE produces a more even background but fails to faithfully represent the faint charge-density-wave (CDW) signal. Combining MAE with MS-SSIM shows a background that is more consistent with the one of the high-count (HC) frame. Additionally, structural patterns including the CDW signal are clearly enhanced.

%\newpage

\subsection{Comparison of CNN-based noise filtering and conventional denoising (smoothing)}

\begin{figure*}
\centering
\includegraphics[width=0.7\textwidth]{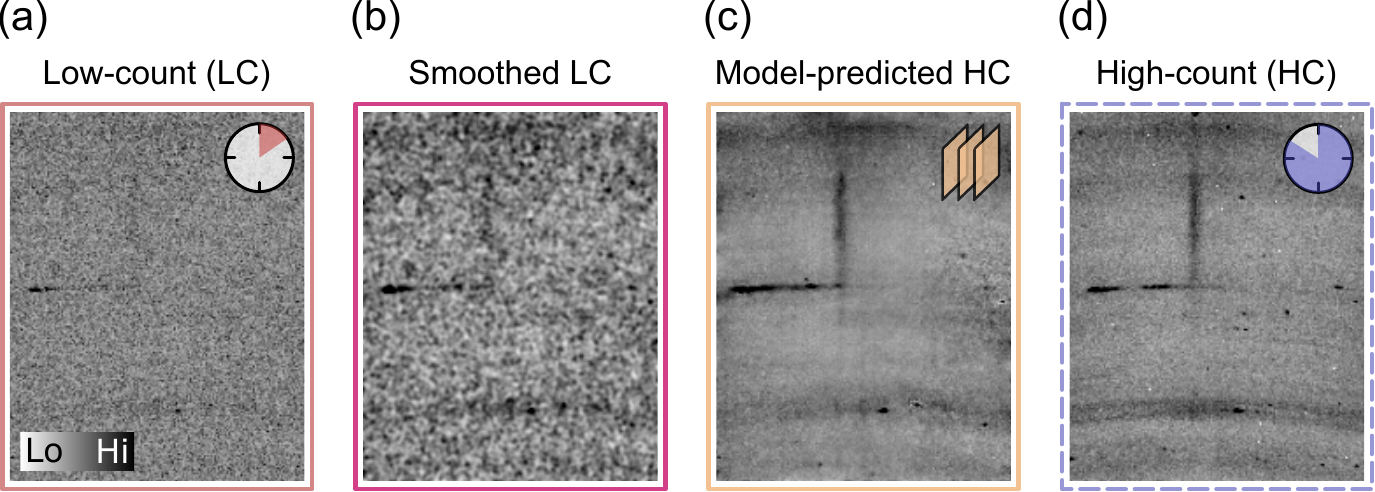}
\caption{Comparison of CNN-based denoising and conventional Gaussian smoothing. (a) Low-count (LC) frame. (b) Gaussian smoothed low-count frame in (a) using a standard deviation of 1. (c) Denoised low-count frame in (a) using a trained CNN. (d) High-count (HC) frame for comparison.}
\label{fig:cnn_vs_smooth}
\end{figure*}

A commonly used practice to reduce the noise in (2D) data is smoothing. One example is Gaussian smoothing where the noisy data is convoluted with a Gaussian kernel of a certain standard deviation. In Figure~\ref{fig:cnn_vs_smooth} we compare a conventional Gaussian smoothing approach and CNN-based noise filtering. While the Gaussian smoothed low-count (LC) frame results in a reduction of the high-frequency noise, it inevitably blurs the data~\cite{KimRSI2021}. On the contrary, the LC frame produced by the trained CNN effectively suppresses the high-frequency noise present in the LC data while the weak CDW signal -- barely visible in the LC frame -- is strongly enhanced. This is achieved by a significant improvement in the local signal continuity following the application of the CNN-based noise filtering.

%\newpage

\subsection{Overfitting}

\begin{figure*}
\centering
\includegraphics[width=0.95\textwidth]{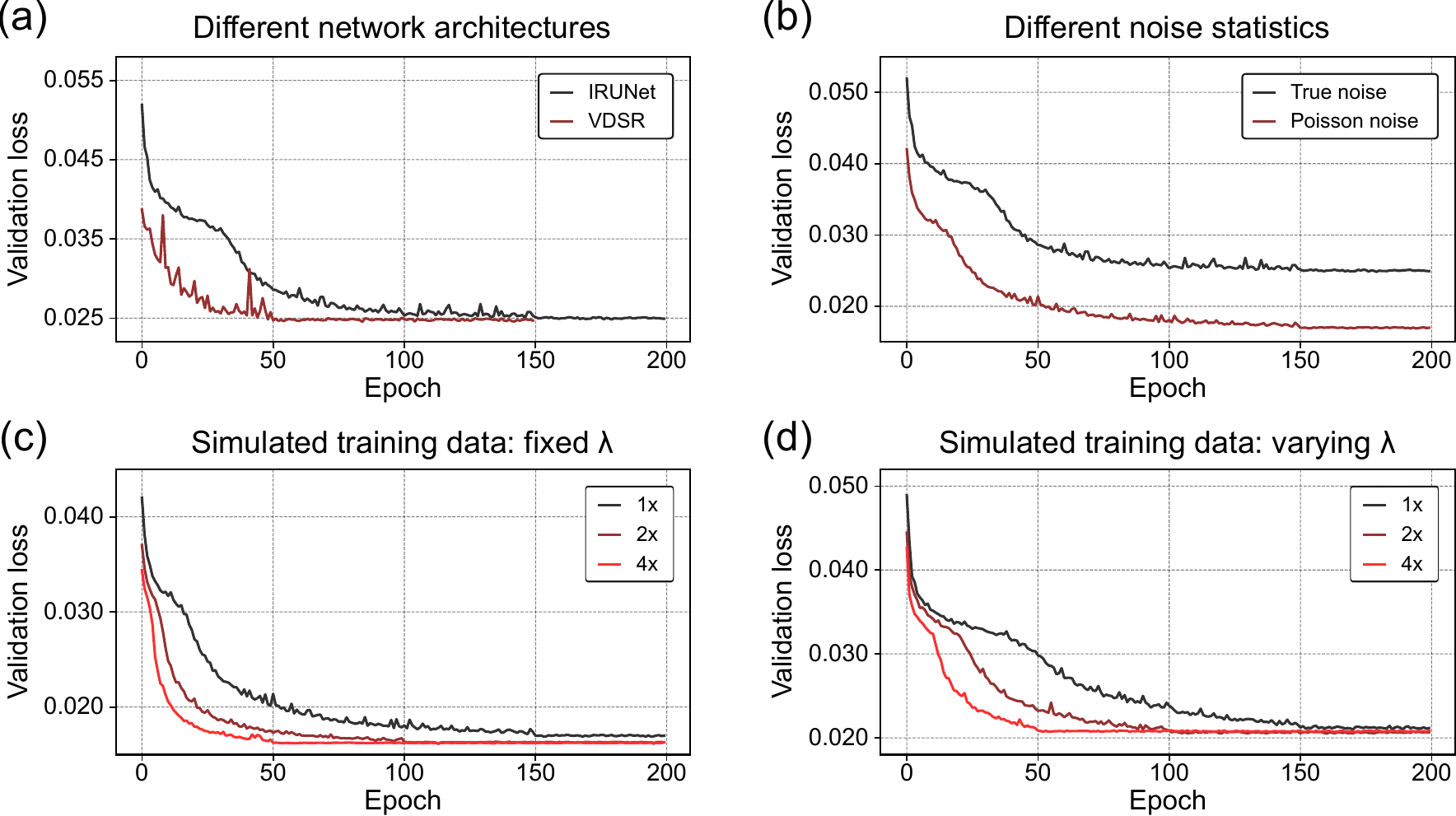}
\caption{Loss curves showing training epoch against validation loss for different training scenarios. (a) Different neural-network architectures. (b) Training on different noise statistics (experimental or simulated Poisson noise). (c) Simulating artificial low-count images with the same Poisson statistics (fixed $\lambda$) for different amounts of artificial training data. For example 2x refers to double the amount with respect to the size of the original training data set. (d) Simulating artificial low-count images with varying $\lambda$ (multiscale training) for different amounts of artificial training data.}
\label{fig:loss_curves}
\end{figure*}

During the training process, we evaluate the neural-network performance using a separate validation data set as mentioned in the main text. We keep track of the resulting validation loss to ensure optimal convergence without overfitting, which shows itself in an increase of the validation loss. %Overfitting is usually a consequence of a too small training dataset. 
It implies that the model cannot perform well on unseen data, as it might have overly specialized in learning the features of the training data. In \mbox{Figure~\ref{fig:loss_curves}(a)} we compare the loss curves for the two used neural network architectures, IRUNet~\cite{gil_zuluaga_blind_2021} and VDSR~\cite{kim_accurate_2016}. In \mbox{Figure~\ref{fig:loss_curves}(b)}, loss curves for different training data statistics are shown. \mbox{Figure~\ref{fig:loss_curves}(c,d)} shows the effect of simulated counting statistics, which will be further elaborated in the next section.

\subsection{More artificial training data, multiscale training, and network modifications}

One of the advantages of artificial training data is the fact that one can simulate an arbitrary amount of low-count (LC) frames given a single high-count (HC) frame by drawing random samples from the underlying Poisson distribution with a fixed $\lambda$ factor as described in the main text. Increasing the amount of training data is a well-known approach to enhance the performance of neural networks. Additionally, the used experimental X-ray diffraction training data consists of LC (HC) pairs with exposure times of 1 (20) seconds for the most part ($>80\%$). However, there are some cases where the frames have been counted for slightly different times, which effectively changes the $\lambda$ factor of the Poisson distribution. In those cases where the statistics differ throughout the test data set, one should benefit from a multiscale (MS) training approach where the network is trained on LC data with different counting times (varying~$\lambda$)~\cite{Zhang2017,KimRSI2021}. Both of these potential improvements regarding the training with artificial training data have been studied. For training a MS model we choose a uniform distribution of 100 $\lambda$ values $\in \left[ 0.001, 0.1 \right]$. The convergence of the validation loss for these training approaches is shown in \mbox{Figure~\ref{fig:loss_curves}(c,d)} where the learning rate has been reduced by half after the first 150, 100, and 50 epochs for 1x, 2x, and 4x amount of training data respectively. We note that the final validation loss does not change that much. However, the amount of required training epochs for convergence is strongly reduced when using more training data. Doubling the amount of training data will roughly half the amount of required epochs. We would like to point out that while the amount of epochs is reduced, the overall training time does in fact not decrease but rather scales with the amount of training data. The training time required for convergence is around 7, 11, and 16 hours for 1x, 2x, and 4x data respectively using an Nvidia Tesla P100 GPU with 10 GB of VRAM. The performance results of mentioned training approaches are summarised in \mbox{Table~\ref{tab:performance_table_ms}} using the IRUNet architecture and 50 different CDW signals as described in the main text. We observe that additional artificial training data with the same statistics (fixed $\lambda$) is still inferior to training with real experimental data. We furthermore observe that employing a multiscale approach does benefit training with artificial noise, sometimes even attaining comparable performance to training with actual experimental data. A multiscale approach could therefore prove valuable when dealing with only a limited amount of experimental training data, provided that the underlying statistics are well-known. %A multiscale approach could therefore be useful in the case when there is only a limited amount of available experimental training data, provided that the underlying statistics are well-known.

%In this work, 
Additionally, we have implemented a VDSR network architecture~\cite{kim_accurate_2016} without the final residual layer. We find that removing said layer doesn't lead to a significant change in performance as shown in \mbox{Table~\ref{tab:performance_table_vdsr_res_vs_no_res_layer_exp_exp}} and~\ref{tab:performance_table_iq_metrics}.
% we show the same results for training and evaluation on experimental data using the VDSR network with and without a final residual layer. %We observe only a marginal improvement when using a residual layer.

% NOTE: I would write that training on aritifical noise using a multiscale procedure might be very helpful in the cases where not enough real experimental data is available. However, if one would have the same diversity of noise statistics with real experimental data then training using real noise should still be the optimal way of training denoising neural networks when the aim is to remove real experimental noise.

%\newpage

\begin{table*}[ht]
\renewcommand{\arraystretch}{1.5}
\centering
\caption{Average Gaussian fitting results of different training scenarios using the IRUNet network architecture. The first column indicates the amount of used (artificial) training data and whether a multiscale (MS) procedure has been utilized. For example, using double the amount of artificial training data and applying a multiscale procedure (2x Poisson MS). The results when training on the original amount of experimental training data is shown at the bottom for comparison (Exp. $\rightarrow$ Exp.) and are additionally highlighted in bold for visual guidance.}
\label{tab:performance_table_ms}

\vspace{0.5cm}

\begin{tabular}{M{0.15}|M{0.085}|M{0.085}|M{0.085}|M{0.085}|M{0.085}|M{0.085}|M{0.085}|M{0.085}|M{0.085}|}
\cline{2-10}
 & $\mu_h$ ($\times 10^2$) & $\mu_k$ ($\times 10^2$) & $\mu_\ell$ ($\times 10$) & $\sigma_h$ ($\times 10^2$) & $\sigma_k$ ($\times 10^2$) & $\sigma_\ell$ ($\times 10$) & SRBR$_h$ & SRBR$_k$ & SRBR$_\ell$\\
 \hline
\multicolumn{1}{|c|}{1x Poisson} & 0.75 (04) & 0.87 (11) & 4.17 (14) & 0.31 (04) & 1.65 (13) & 1.37 (17) & 1.02 (25) & 0.95 (13) & 0.97 (04)\\
\multicolumn{1}{|c|}{2x Poisson} & 0.66 (35) & 1.68 (14) & 3.11 (13) & 0.39 (35) & 1.53 (11) & 1.55 (15) & 0.87 (15) & 1.04 (34) & 0.95 (04)\\
\multicolumn{1}{|c|}{4x Poisson} & 0.44 (02) & 2.00 (10) & 3.24 (15) & 0.16 (02) & 0.94 (15) & 1.28 (17) & 0.88 (08) & 0.98 (06) & 1.04 (04)\\
\multicolumn{1}{|c|}{1x Poisson MS} & 0.32 (07) & 0.73 (07) & 1.23 (12) & 0.33 (07) & 0.80 (08) & 1.10 (15) & 0.92 (13) & 1.13 (02) & 1.11 (04)\\
\multicolumn{1}{|c|}{2x Poisson MS} & 0.35 (02) & 0.69 (07) & 1.08 (12) & 0.18 (02) & 0.66 (07) & 1.01 (15) & 0.89 (08) & 1.06 (02) & 1.09 (04)\\
\multicolumn{1}{|c|}{4x Poisson MS} & 0.45 (02) & 0.65 (07) & 2.66 (13) & 0.19 (02) & 0.72 (07) & 1.22 (16) & 0.91 (07) & 1.06 (02) & 1.15 (05)\\
\multicolumn{1}{|c|}{Exp. $\rightarrow$ Exp.} & \textbf{0.19 (03)} & \textbf{0.65 (07)} & \textbf{1.47 (12)} & \textbf{0.31 (03)} & \textbf{0.69 (08)} & \textbf{1.50 (15)} & \textbf{1.41 (24)} & \textbf{1.00 (02)} & \textbf{1.21 (05)}\\ % #4
% \multicolumn{1}{|c|}{1x Exp.} & \textbf{0.30 (03)} & \textbf{0.64 (08)} & \textbf{1.37 (12)} & \textbf{0.20 (02)} & \textbf{0.88 (08)} & \textbf{1.31 (14)} & \textbf{0.85 (08)} & \textbf{1.09 (02)} & \textbf{1.03 (04)}\\ % #2
\hline
\end{tabular}
\end{table*}

\begin{table*}[ht]
\renewcommand{\arraystretch}{1.5}
\centering
\caption{Average Gaussian fitting results for the VDSR network architecture with ($\oplus$) and without a final residual layer, trained and evaluated on experimental data.}
\label{tab:performance_table_vdsr_res_vs_no_res_layer_exp_exp}

\vspace{0.5cm}

\begin{tabular}{M{0.15}|M{0.085}|M{0.085}|M{0.085}|M{0.085}|M{0.085}|M{0.085}|M{0.085}|M{0.085}|M{0.085}|}
\cline{2-10}
 & $\mu_h$ ($\times 10^2$) & $\mu_k$ ($\times 10^2$) & $\mu_\ell$ ($\times 10$) & $\sigma_h$ ($\times 10^2$) & $\sigma_k$ ($\times 10^2$) & $\sigma_\ell$ ($\times 10$) & SRBR$_h$ & SRBR$_k$ & SRBR$_\ell$\\
 \hline
 \multicolumn{1}{|c|}{VDSR} & 0.32 (02) & 0.63 (08) & 1.11 (11) & 0.16 (02) & 0.73 (08) & 0.95 (14) & 0.97 (09) & 1.09 (02) & 1.25 (04)\\
\multicolumn{1}{|c|}{VDSR ($\oplus$)} & 0.20 (10) & 0.63 (07) & 1.31 (01) & 0.14 (01) & 0.64 (08) & 0.81 (14) & 1.00 (11) & 1.13 (02) & 1.47 (05)\\
\hline
\end{tabular}
\end{table*}

% \begin{table*}
% \renewcommand{\arraystretch}{1.5}
% \centering
% \caption{Average Gaussian fitting results for the VDSR network architecture with ($\oplus$) and without a residual layer, trained on artificial and evaluated on experimental data.}
% \label{tab:performance_table_vdsr_res_vs_no_res_layer_poisson_exp}

% \vspace{0.5cm}

% \begin{tabular}{M{0.15}|M{0.085}|M{0.085}|M{0.085}|M{0.085}|M{0.085}|M{0.085}|M{0.085}|M{0.085}|M{0.085}|}
% \cline{2-10}
%  & $\mu_h$ ($\times 10^2$) & $\mu_k$ ($\times 10^2$) & $\mu_\ell$ ($\times 10$) & $\sigma_h$ ($\times 10^2$) & $\sigma_k$ ($\times 10^2$) & $\sigma_\ell$ ($\times 10$) & SBR$_h$ & SBR$_k$ & SBR$_\ell$\\
%  \hline
% \multicolumn{1}{|c|}{VDSR} & 0.46 (02) & 0.72 (14) & 2.62 (11) & 0.19 (02) & 1.28 (18) & 1.18 (14) & 0.94 (07) & 1.15 (03) & 1.25 (04)\\
% \multicolumn{1}{|c|}{VDSR ($\oplus$)} & 1.81 (06) & 11.5 (07) & 1.31 (01) & 0.14 (01) & 0.64 (08) & 0.81 (14) & 1.00 (11) & 1.13 (02) & 1.47 (05)\\
% \hline
% \end{tabular}

% \end{table*}

%\newpage

\subsection{Evaluation using standard image quality metrics}

In this study, we assessed the denoising performance of the trained neural networks based on physical signal properties, including the signal-to-residual-background ratio (SRBR) of the CDW peak. Such an evaluation removes potential ambiguities that might emerge when using standard image quality metrics, such as peak signal-to-noise ratio (PSNR) or structural similiarty (SSIM)~\cite{wang_image_2004,wang_multiscale_2003}, as these metrics necessitate a noise-free ground truth image. However, given the nature of experimental data, a finite amount of noise persists even with high counting statistics. This inherent noise renders an evaluation purely based on mentioned image quality metrics less favorable. For completeness, those metrics are summarized in \mbox{Table~\ref{tab:performance_table_iq_metrics}} next to unambiguous mean absolute (MAE) and mean squared errors (MSE).

\begin{table*}[ht]
\renewcommand{\arraystretch}{1.5}
\centering
\caption{Denoising performance using standard image quality metrics for different neural-network architectures and training scenarios. The evaluation has been performed on true experimental data from the separate test set mentioned in the main text. The values are given as the mean (median) over all test images. The first column indicates the used network architecture. The second column refers to the amount of used (artificial) training data and whether a multiscale (MS) procedure has been utilized. For example, using double the amount of artificial training data and applying a multiscale procedure (2x Poisson MS). The results when training on the original amount of experimental training data is shown at the bottom of each network section for comparison (Exp. $\rightarrow$ Exp.) and are additionally highlighted in bold for visual guidance.}
\label{tab:performance_table_iq_metrics}

\vspace{0.5cm}

\begin{tabular}{|M{0.1}|M{0.15}|M{0.15}|M{0.15}|M{0.15}|M{0.15}|}
\cline{3-6}
\multicolumn{1}{c}{} & & MAE - L1 ($\times 10$) & MSE - L2 ($\times 10^2$) & PSNR (dB) & MS-SSIM\\
 \hline
 \multirow{7}{*}{IRUNet} & \multicolumn{1}{c|}{1x Poisson} & 0.153 (0.109) & 0.073 (0.021) & 35.510 (36.838) & 0.952 (0.977)\\
& \multicolumn{1}{c|}{2x Poisson} & 0.141 (0.108) & 0.059 (0.020) & 35.766 (36.995) & 0.953 (0.978)\\
& \multicolumn{1}{c|}{4x Poisson} & 0.138 (0.107) & 0.049 (0.020) & 35.871 (36.999) & 0.962 (0.978)\\
& \multicolumn{1}{c|}{1x Poisson MS} & 0.122 (0.118) & 0.027 (0.024) & 35.819 (36.178) & 0.974 (0.977)\\
& \multicolumn{1}{c|}{2x Poisson MS} & 0.115 (0.112) & 0.025 (0.022) & 36.259 (36.638) & 0.975 (0.978)\\
& \multicolumn{1}{c|}{4x Poisson MS} & 0.117 (0.113) & 0.025 (0.022) & 36.185 (36.564) & 0.975 (0.977)\\
& \multicolumn{1}{c|}{Exp. $\rightarrow$ Exp.} & \textbf{0.118 (0.111)} & \textbf{0.028 (0.022)} & \textbf{36.00 (36.530)} & \textbf{0.973 (0.977)}\\
\hline
\multirow{2}{*}{VDSR} & \multicolumn{1}{c|}{1x Poisson} & 0.123 (0.114) & 0.032 (0.022) & 35.744 (36.501) & 0.965 (0.975)\\
& \multicolumn{1}{c|}{Exp. $\rightarrow$ Exp.} & \textbf{0.116 (0.111)} & \textbf{0.027 (0.022)} & \textbf{36.011 (36.506)} & \textbf{0.973 (0.976)}\\
\hline
\multirow{2}{*}{VDSR ($\oplus$)} & \multicolumn{1}{c|}{1x Poisson} & 0.127 (0.114) & 0.034 (0.022) & 35.794 (36.544) & 0.965 (0.976)\\
& \multicolumn{1}{c|}{Exp. $\rightarrow$ Exp.} & 0.127 (0.121) & 0.030 (0.025) & 35.540 (35.973) & 0.971 (0.973)\\
\hline
\end{tabular}

\end{table*}

%\newpage

\subsection{Influence of training data shuffling, random seed, and optimizer}

% The results described in the following are obtained for the IRUNet network architecture trained with the same hyperparameters as described in the method section of the main text. After training the network was additionally evaluated using standard image quality metrics described previously.

\begin{figure*}
\centering
\includegraphics[width=0.95\textwidth]{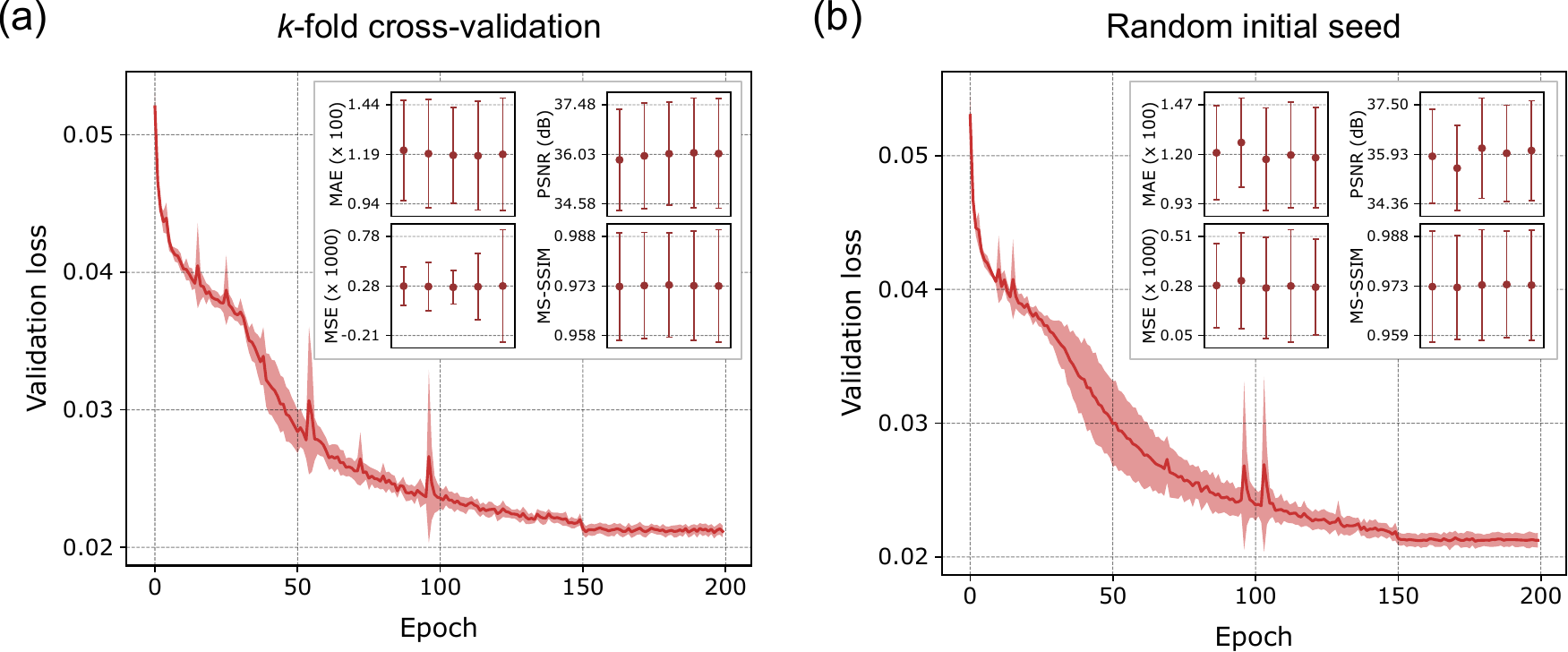}
\caption{Denoising performance for (a) $k$-fold cross-validation ($k=5$) and (b) different random seeds (5) for the initialization of the network weights using the IRUNet architecture. The validation loss curve corresponds to the mean, while the shaded area corresponds to the standard deviation of the validation loss over the performed training runs. The insets in (a) and (b) show mean values (dots) and standard deviation (error bars) of standard image quality metrics obtained after evaluating the trained networks on the separate test set described in the main text.}
\label{fig:randomization}
\end{figure*}

A common technique to evaluate the performance of a trained machine-learning model is $k$-fold cross-validation~\cite{raschka_model_2020} where the entire data set is split into $k$ equally-sized parts. One part is used for validation while the remaining $k-1$ parts are combined into the training data set. As described in the main text, we use a 4:1 splitting ratio for training and validation data set (3280 training and 820 validation pairs). In \mbox{Figure ~\ref{fig:randomization}(a)} we show the result of cross-validation for $k=5$ splits. In particular, we observe that the final loss and denoising performance is independent of the chosen training-validation splitting. 

In general, the chosen random seed, used for the initialization of the network parameters, can significantly influence the final performance when training deep-learning models~\cite{alahmari_challenges_2020,picard_torchmanual_seed3407_2023}. It is thus advised to verify that training results are reproducible. As such, we conducted an experiment where we varied the random seed -- see \mbox{Figure~\ref{fig:randomization}(b)}. While we observe slightly larger fluctuations of the loss curve and image quality metrics compared to \mbox{Figure~\ref{fig:randomization}(a)}, the final performance does not appear to be strongly affected.

\begin{figure*}
\centering
\includegraphics[width=0.95\textwidth]{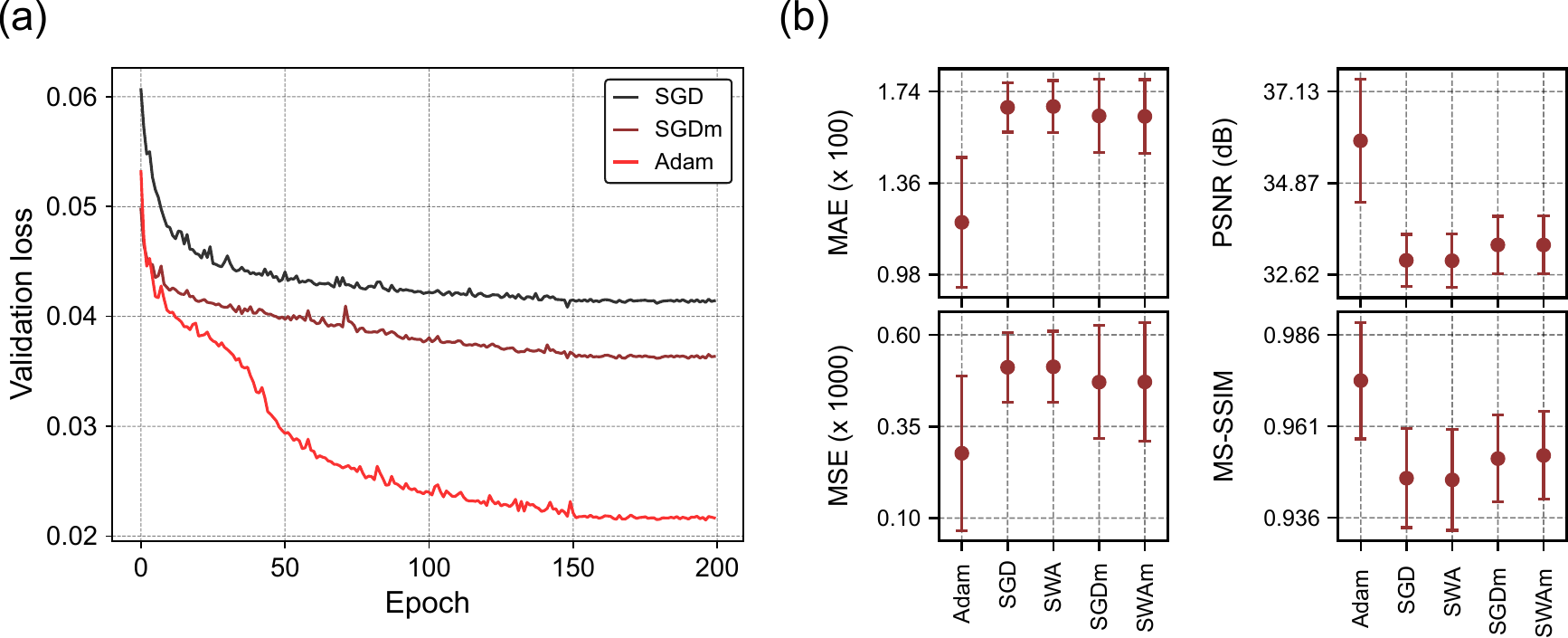}
\caption{Denoising performance for different optimizers using the IRUNet architecture. Next to adaptive momentum estimation (Adam), gradient descent methods such as stochastic gradient descent (SGD), and stochastic weight averaging (SWA), both with and without momentum, have been considered. (a) Validation loss curves for Adam and SGD with (m) and without momentum. (b) Mean values (dots) and standard deviation (error bars) of standard image quality metrics obtained after evaluating the trained networks on the separate test set described in the main text.}
\label{fig:optimizer}
\end{figure*}

Finally, an adaptive momentum estimation (Adam) optimizer~\cite{kingma_adam_2017} was used for training the deep neural networks in this work. However, it has been reported that stochastic gradient descent (SGD) methods are better at generalizing and finding a broader global optimum~\cite{wilson_marginal_2018,zhou_towards_2021}. In \mbox{Figure~\ref{fig:optimizer}(a)} we compare loss curves of Adam, SGD, and stochastic weight averaging (SWA)~\cite{izmailov_averaging_2019} while in \mbox{Figure~\ref{fig:optimizer}(b)} we show their denoising performance using standard image quality metrics. For SGD and SWA, variants with and without momentum have been considered. As described in the method section of the main text, a learning rate of $5 \times 10^{-4}$ has been used for Adam while SGD and SWA runs were performed using a learning rate of $1 \times 10^{-1}$. These learning rates have been found to yield the best final validation loss for the respective optimizers over 200 epochs using the IRUNet architecture. For the last 50 epochs the learning rate has been reduced by half. A momentum of 0.9 has been chosen for the momentum-variants of SGD~(SGDm) and SWA~(SWAm). Overall, we find that Adam results in a considerably better denoising performance compared to SGD and SWA despite the fact that it has a larger spread (error bars in Figure~\ref{fig:optimizer}(b)). We also observe that SGD (SWA) with momentum tend to yield slightly better results compared to their non-momentum counterparts.

% \newpage

\subsection{Receptive field of the neural networks}

A key property of every convolutional neural network (CNN) is its receptive field, which defines the amount of context information available to the neural network for learning. Ideally, the receptive field should be large enough to capture the features of interest. In our case, these are rod-shaped CDW order that have a spatial extension of up to a third of the larger image dimension, roughly 80 pixels. For a simple single-path network such as VDSR, the receptive field using $D$ consecutive convolutional layers with kernel size 3 and unit stride is given as (2$D$+1)$\times$(2$D$+1)~\cite{kim_accurate_2016,Zhang2017,araujo2019computing}. %\sout{In our case} 
In this work, we used 20 convolutional layers resulting in a receptive field of 41$\times$41 pixels. For more complicated network structures, such as IRUNet, the receptive field cannot be calculated in a straight-forward fashion~\cite{araujo2019computing,luo_understanding_2017} because it strongly depends on whether, for example, skip connections, non-linear activation functions, and pooling layers are utilized. Using a gradient-based backpropagation method~\cite{noauthor_receptivefield_2023} we estimate the receptive field of the used IRUNet network to be around 170$\times$200 pixels, which is much larger than the receptive field of VDSR. Nevertheless, IRUNet does not yield superior results compared to VDSR, suggesting that a larger receptive field does not necessarily relate to a better denoising performance in this context.

\begin{figure*}
\centering
\includegraphics[width=0.6\textwidth]{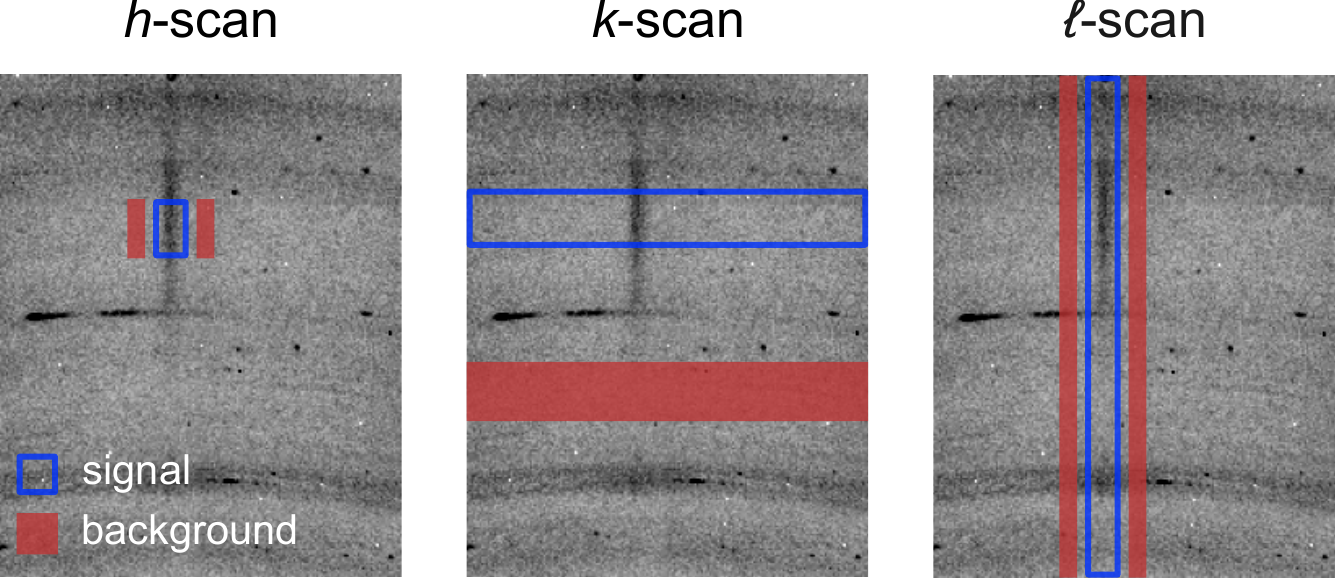}
\caption{Placement of signal and background region-of-interest (ROI) for $h$, $k$, and $\ell$ scans. In the case of $h$ and $\ell$ scans, the background ROI consists of two rectangles of equal sizes, situated next to the signal ROI. The combined size of these two rectangles matches the size of the signal ROI. For $k$ scans, a single rectangle is used as the background instead.}
\label{fig:background_subtraction}
\end{figure*}

\subsection{Background subtraction}

As described in the main text, a background subtraction has been performed prior to the line-profile analysis of the charge-density-wave signal. This process involves the summation of pixel intensities within a region-of-interest (ROI) around the signal and subtraction of neighbouring background ROIs. The placement of signal and background ROIs for individual $h$, $k$, and $\ell$ scans is illustrated in \mbox{Figure~\ref{fig:background_subtraction}}.

\end{document}